\documentstyle[12pt]{article}
\def\beq{\begin{equation}}
\def\eeq{\end{equation}}
\def\bea{\begin{eqnarray}}
\def\eea{\end{eqnarray}}

\begin {document}
\begin{titlepage}
September 1993 \hfill DESY 93-??? \\
\begin{flushright}
[ HU Berlin-IEP-93/4 ] \\
\end{flushright}
\mbox{ }  \hfill hepth@xxx/9309079
\vspace{6ex}
\Large
\begin {center}
\bf{Operator product expansion of the energy momentum tensor in 2D conformal
field theories on manifolds with boundary}
\end {center}
\large
\vspace{3ex}
\begin{center}
H. Dorn  \footnote{e-mail: dorn@ifh.de } and V. Preu\ss
\end{center}
\normalsize
\it
\vspace{3ex}
\begin{center}
Fachbereich Physik der Humboldt--Universit\"at \\
Institut f\"ur Elementarteilchenphysik \\
Invalidenstra\ss e 110, D-10099 Berlin, Germany
\end{center}
\vspace{6 ex }
\rm
\begin{center}
\bf{Abstract}
\end{center}
\vspace{3ex}
Starting from the well-known expression for the trace anomaly we derive the
$T\cdot T$ operator product expansion of the energy-momentum tensor in 2D
conformal theories defined in the upper halfplane $without$ making use of the
additional condition of no energy-momentum flux across the boundary. The OPE
turns out to be the same as in the absence of the boundary. For this result
it is crucial that the trace anomaly is proportional to the Gau\ss-Bonnet
density. Some relations to the $\sigma$ - model approach for open strings
are discussed.
\end {titlepage}
\newpage
\setcounter{page}{1}
\pagestyle{plain}
\section {Introduction}
In a conformal field theory defined on the two dimensional infinite plane
the operator product expansion (OPE) of the energy-momentum tensor
\beq
T_{++}(z)T_{++}(w)~=~\frac{\frac{c}{2}}{(z-w)^{4}}+\frac{2T_{++}(w)}{
(z-w)^{2}}+\frac{\partial_{+} T_{++}(w)}{z-w}~+~fin.~terms
\label{1}
\eeq
plays a very fundamental role. There is a similar formula for $T_{--}\cdot
T_{--}
$ obtained by $z,w \rightarrow \bar{z},\bar{w} $. The trace $T_{+-}$ as well as
$T_{++} \cdot T_{--} $ vanish up to contact terms. Introducing moments of
$T_{++}$ and $T_{--} $ one gets from (\ref{1}) two independent Virasoro
algebras
with the same central charge $c$. A further textbook fact relates the central
charge to the Weyl anomaly of the theory generalized to curved 2D manifolds
\beq
T_{a}^{a}(z)~=~\frac{c}{48\pi} R(z)~~.
\label{2}
\eeq
Of course the trace vanishes in the flat limit $R\rightarrow 0$. \\

For manifolds with boundary there is no flux of energy-momentum across the
boundary if there are no additional interactions localized at the boundary.
In the following we concentrate ourselves to the simplest case, the upper
half-plane. Then we have $T_{12}=0$ on the real axis, i.e.
\beq
T_{++}(z)~=~T_{--}(z)~~,~~~if~ Im(z)=0~~.
\label{3}
\eeq
The more general situation of present boundary interactions is relevant for
instance in the $\sigma $-model picture of open strings
\cite{FrTs,Ts2,DO,Callan} as well as for surface critical
behaviour in statistical mechanics \cite{Diehl}. Away from criticality,
i.e. conformal invariance, energy-momentum flux across the boundary is
allowed, of course. However, in the conformal limit the conserved current
$j_{m}=\epsilon ^{n}T_{mn} $ for boundary conserving conformal transformations
$\epsilon $
has to be parallel to the boundary. This again forbids energy-momentum flux
across the boundary. The consequences of (\ref{3}) are well-known \cite{C1}.
Using analyticity one can continue into the lower half-plane
via $T_{--}(z)=T_{++}(\bar{z}) $. Then there is only one independent component
of the energy-momentum tensor and only one Virasoro algebra. The OPE has the
same form (\ref{1}) as in the absence of the boundary. The net result of this
repetition can be summarized as follows: To find the conformal cases out of the
generalized $\sigma $-models defined in the upper half-plane one has to look
for solutions of $T_{a}^{a}~=~0,~~Re(z)\geq0 ~and~T_{++}~=~T_{--},~~Im(z)=0
$. \\

Our paper will be a contribution towards the elimination of the condition
(\ref{3})
as an independent one. Although we are not showing that $T_{a}^{a}~=~0$
implies (\ref{3}), we derive (\ref{1}) for the half-plane
from the knowledge of the trace $T_{a}^{a}$ alone. As mentioned already, in
the standard derivation of (\ref{1}) condition (\ref{3}) is used as an
essential input. Hence in some sense our note is in the tradition of the
Curci-Paffuti theorem \cite{CP,Ts1,Osb2} which uncovers
relations between otherwise independent coefficient functions in the trace
of the energy-momentum tensor. Vanishing of the $\bar{\beta}_{i}$-functions for
$i\neq \phi $ ($\phi$ dilaton) results in $\bar{\beta}_{\phi}=const$. The
constant value of $\bar{\beta }_{\phi}$ plays the role of the central charge
in the corresponding conformal theory. The theorem has been extended to the
case
of $\sigma $-models on manifolds with boundary \cite{BD1,Osb1}.
The vanishing of all non-dilaton bulk as well as boundary
$\bar{\beta}$-functions
results in constant $\bar{\beta }_{\phi}$ and $\bar{\beta}_{\hat{\phi}}$ (
$\hat{\phi} $ dilaton coupling to the boundary). Using Wess-Zumino type
consistency relations one can show \cite{Osb1} that the constant values of
$\bar{\beta}_{\phi}$ and $\bar{\beta}_{\hat{\phi}}$ are correlated. The trace
$T_{a}^{a}$  for the extension to curved 2D manifolds has to be proportional
to the Gau\ss-Bonnet density (see also ref. \cite{DW})
\beq
T_{a}^{a}(z)~=~\frac{c}{48 \pi} R(z)~+~\frac{\hat{c}}{24 \pi} \int ds~
\delta ^{(2)}(z-z(s))~k(z(s))
\label{4}
\eeq
with
\beq
c~~=~~\hat{c}~~.
\label{4a}
\eeq
$k(z(s))$ is the geodesic curvature of the boundary at the point $z(s)$ and $s$
the length parameter measured with the 2D metric $g_{ab}$.\\

Starting from (\ref{4}) we will derive the $T\cdot T$ - OPE by adapting the
technique of ref. \cite{EO} to the boundary case. The calculations will be
done for general uncorrelated $c,~\hat{c}$. At the end we will get (\ref{1})
only for $c=\hat{c}$. As a by-product of this analysis we find, turning the
argument around, that insisting on (\ref{3}) as an independent input yields an
alternative proof of $c=\hat{c}$.

\section {Derivation of the OPE}
We start from the conformal Ward identity \cite {EO} with the Weyl anomaly
adapted in the sense of (\ref {4}) to the case of a manifold with boundary
\bea
\frac{1}{2} \int d^{2}z \sqrt {g}(P \epsilon )_{ab} \langle T^{ab}(z)~\Phi _{1}
(z_{1})...\Phi _{N} (z_{N}) \rangle&=& \nonumber \\
=~\Big(\frac {c}{48\pi }\int d^{2}z \sqrt {g} R \nabla _{a} \epsilon ^{a}(z)&
+&\frac {\hat c}{24\pi}\oint k(z(s)) \nabla _{a} \epsilon ^{a}ds\Big) \langle
\prod
_{j} \Phi _{j} \rangle \nonumber \\
-~\sum _{k=1}^{N} \Big(\epsilon ^{a}(z_{k})\frac {\partial}{\partial z_{k}
^{a}}~+~
\frac {d_{k}}{2} \nabla _{a} \epsilon ^{a} (z_{k})&+&is_{k}(\frac {\epsilon
_{ab}
}{2} \nabla ^{a} \epsilon ^{b} + \omega _{a} \epsilon ^{a} (z_{k}))\Big)
\langle \prod _{j} \Phi _{j} \rangle.
\label {5}
\eea
$\epsilon ^{a}(z)$ is a diffeomorphism with $P\epsilon$ defined by
\beq
(P\epsilon)_{ab}~=~\nabla _{a}\epsilon _{b}~+~\nabla _{b}\epsilon _{a}~-~
g_{ab}\nabla _{c}\epsilon ^{c}~~,
\label{6}
\eeq
while $d_{k} ,~s_{k}$ are the dimensions and spins of the primary fields
$\Phi _{k}(z_{k})$. The identity (\ref{5}) will be functionally differentiated
with respect to $g_{mn}(w)$. This yields, among other terms, a
double insertion of the energy-momentum tensor. By a suitable choice of $
\epsilon (z)$ one can localize the expressions finally. This standard procedure
\cite{FMS,EO} in our case gets two new aspects. The diffeomorphism has to
respect the boundary and one has to handle the metric variation of the
geodesic curvature $k$ as well as the length parameter $s$ on the boundary.\\

The energy-momentum tensor in (\ref{5}) is defined by
\beq
T^{ab}~=~ \frac {2}{\sqrt {g}}~ \frac {\delta S}{\delta g_{ab}}~~.
\label{7}
\eeq
Hence the natural position of indices is the upper one for $T$ and $\epsilon$
but the lower one for $g,~\partial $ and $\nabla $. This means
\bea
\delta \Big( \sqrt {g} (P\epsilon )_{ab} \Big)&=&\sqrt{g}\Big[ \nabla _{d} (
\epsilon ^{d} \delta g_{ab}-\frac{1}{2} \epsilon ^{d} g_{ab} g^{mn} \delta
g_{mn}) \nonumber \\
&+& \frac{1}{2} g^{mn} \delta g_{mn} (\nabla _{a} \epsilon _{b}+\nabla _{b}
\epsilon _{a}) \nonumber \\
&+& \delta g_{bd} \nabla _{a} \epsilon ^{d}+ \delta g_{ad} \nabla _{b}
\epsilon ^{d}- 2 \delta g_{ab} \nabla _{d} \epsilon ^{d} \Big]~~.
\label{8}
\eea
Therefore, the flat limit of the derivative of the l.h.s. of (\ref{5})
with respect to $g_{--}(w)$ is
\bea
\frac{\delta (l.h.s.~of (\ref{5}))}{\delta g_{--}(w)}\Big{\vert}_{flat}~=~
\Big{\langle} \Big(
&4& \int d^{2}z~ T_{++}(w)[\partial _{-} \epsilon ^{+}(z)T_{++}(z) +
\partial _{+} \epsilon ^{-}(z) T_{--}(z)] \nonumber \\
&+&i~\int d z^{1}\delta ^{(2)} (z-w) (\epsilon ^{+}(z)-\epsilon ^ {-}(z))
T_{++} (z) \nonumber \\
+~4\partial _{+} \epsilon ^{-} (w) T_{+-} (w)
&-&4\partial _{+} \epsilon ^{+} (w) T_{++} (w)
{}~-~2 \epsilon ^{d} (w) \partial _{d} T_{++} (w) \Big) \prod _{j}\Phi _{j}
\Big{\rangle} ~.
\label{9}
\eea
We use the notation $z=z^{+}=z^{1}+iz^{2},~~\bar{z}=z^{-}=z^{1}-iz^{2}$.
The integral over $z^{1}$ has to be taken along the real axis.\\

To differentiate the r.h.s. of (\ref{5}) we first note, that the term under
the sum does not contribute to singularities at $z=w$. Hence we need only
\bea
\delta (\nabla _{a} \epsilon ^{a} )&=&\frac{1}{2} \epsilon ^{c} g^{ab}
\nabla _{c} \delta g_{ab} \\
\label{10}
\delta (\sqrt {g}R)&=& \sqrt {g} (\nabla ^{a} \nabla ^{b} \delta g_{ab}~-~
g^{ab} \nabla ^{2} \delta g _{ab})
\label{11}
\eea
and the variations of $k$ and $s$. The variation of $s$ is trivial. From
$ds^{2}~=~g_{ab} dz^{a}dz^{b}$ we get ($t$ denotes an arbitrary parameter
for the boundary)
\beq
\delta (ds)~=~ \frac{1}{2 \vert \dot{z} \vert } \delta g_{ab} \dot{z}^{a}
\dot{z}^{b}dt~~. \\
\label{12}
\eeq
For $k$ we start from ($'$denotes differentation with respect to $s$)
\beq
k^{2}~=~\nabla (s) z'^{a} \nabla (s) z'^{b} g_{ab}~=~g_{ab}(z''^{a}+z'^{m}
z'^{n} \Gamma _{mn}^{a})(z''^{b}+z'^{j}z'^{l} \Gamma _{jl}^{b})~~.
\\
\label{13}
\eeq
The sign in $\delta k~=~\pm \frac {1}{2} \frac{\delta k^{2}}{k}$ we fix at the
end by comparison with the well-known expression for $k$ in conformal gauge.
To figure  out correctly the variation of $z'$ and $z''$ one has to keep in
mind $\frac{d}{ds}~=~(\dot{z}^{a} \dot{z}^{b} g_{ab})^{- \frac{1}{2}} \frac{d}
{dt}$. With some algebra we find
\beq
(\delta k)_{flat}~=~-k_{flat}(z'^{a}z'^{b}-\frac{1}{2}n^{a}n^{b}) \delta g_{ab}
{}~+~ n^{a}z'^{b}z'^{c} \partial _{b} \delta g_{ac}~-~\frac{1}{2}z'^{a}z'^{b}
\partial _{n} \delta g_{ab}~~.\\
\label{14}
\eeq
After variation the flat limit has been taken. $n$ is the inward unit
normal vector, $\partial _{n}$ denotes
differentiation in the direction of $n$. $k_{flat}$ is the curvature of the
boundary measured with $g_{ab}=\delta _{ab}$.\\

Specializing to the half plane geometry eqs. (\ref{10})-(\ref{12}), (\ref{14})
lead to
\bea
\frac{\delta (r.h.s.~of~(\ref{5}))}{\delta g_{--}(w)} \Big{\vert}_{flat}&=&
\Big{\langle}\prod _{j} \Phi_{j}\Big( \frac{c}{12\pi} \int d^{2}z \partial _{+}
\partial _{+}
\delta ^{(2)}(z-w) \partial _{a} \epsilon ^{a}(z) \nonumber\\
&-& \frac{\hat{c}}{48\pi}\int dz^{1} \partial _{a} \epsilon ^{a}(z)~(
2i\partial _{1}~+~\partial _{2}) \delta ^{(2)}(z-w) \Big) \Big{\rangle}
\nonumber\\
&+& terms~ irrelevant~ for~ the~ OPE~ of~T\cdot T~,
\nonumber
\eea
which after partial integrations with careful book-keeping of boundary terms
yields
\bea
\frac{\delta (r.h.s.~of~(\ref{5}))}{\delta g_{--}(w)} \Big{\vert}_{flat}&=&
\Big{\langle}\prod _{j} \Phi_{j}\Big( \frac{c-\hat{c}}{48\pi} \int dz^{1}
[\partial _{a} \epsilon ^{a}(z) \partial _{2} \delta ^{(2)}(z-w) \nonumber \\
&&~~~~~~~~~~~~~~~~~~~~~~~~~~~-2i\delta ^{(2)}(z-w) \partial _{1} \partial _{a}
\epsilon ^{a}(z)]\nonumber\\
&-& \frac{c}{48\pi}\int dz^{1} \delta ^{(2)}(z-w)\partial_{2} \partial _{a}
\epsilon ^{a}(z)+ \frac{c}{12\pi}~ \partial _{+} \partial _{+} \partial _{a}
\epsilon ^{a}(w) \Big)\Big{\rangle}.
\label{15}
\eea
\\

In the case without boundary $\epsilon ^{+}(z)=\frac{1}{z-v}$ and $\epsilon
^{-}=0$, due to $\partial _{-} \frac{1}{z-v}=\pi \delta ^{(2)}(z-v)$, yields a
localized $T\cdot T$ product built out of $T_{++}T_{++}$ alone \cite{EO}. Since
our diffeomorphism has to keep invariant the real axis we must require
\beq
\epsilon ^{+}(z)~=~\epsilon ^{-}(z)~~if~~z=\bar{z}~~.
\label{16}
\eeq
Therefore, the best we can do is to choose
\beq
\epsilon ^{+}(z)~=~\frac{1}{z-v}~+~\frac{1}{z-\bar{v}}~~;~~~~~
\epsilon ^{-}(z)~=~\frac{1}{\bar{z} -\bar{v}}~+~\frac{1}{\bar{z}-v}~~.
\label{17}
\eeq
This leads to
\beq
\partial _{-} \epsilon ^{+}(z)~=~ \pi ~\delta ^{(2)}(z-v)~=~\partial _{+}
\epsilon ^{-}(z)~~.
\label{18}
\eeq
\\

At this place we should make a comment on the distributions arising as
part of the derivatives of the diffeomorphism. Since $\epsilon ^{a}(z)$
in eq. (\ref{5}) has to be defined globally, this Ward identity as it stands is
valid for regularized versions of (\ref{17}) only. The simplest variant is
\beq
\epsilon ^{+} (z)~=~ \frac{\bar{z} -\bar{v}}{\vert z-v \vert^{2}+ \delta ^{2}}
{}~+~\frac{\bar{z}-v}{\vert z-\bar{v} \vert ^{2}+ \delta ^{2}}~~,~~~~~
\epsilon^{-}(z)~=~\epsilon ^{+}(\bar{z})~~.
\label{19}
\eeq
Only afterwards one can study the limit $\delta \rightarrow 0$. This remark is
helpful in making sense out of some questionable expressions one gets in
calculating with (\ref{17}) directly. For instance, on the formal level instead
of (\ref{18}) one finds $\partial _{-} \epsilon ^{+}(z)=\pi \delta ^{(2)}(z-v)
+\pi
\delta ^{(2)}(z-\bar{v})$. For $v\neq \bar{v}$ the second term can be dropped,
for $v=\bar{v}$ it seems to double the coefficient in front of $\delta ^{(2)}
(z-v)$. But for $v=\bar{v}$ the $z$-integration in the vicinity of $v$ is
restricted to a halfcircle. These two effects conspire in such a way
that the $\delta \rightarrow 0$ limit of $\partial _{-} \epsilon ^{+}$
is indeed (\ref{18}) with $\delta ^{(2)}(z-v)$ understood as the $\delta
$-function of the halfplane, i.e. $\int d^{2}z \delta ^{(2)}(z-v)f(z)~=~f(v)$
for all $v$ with $Im(v)\geq 0$.\\

By going back to the regularized form $\delta >0$ one can proof that the
second line of the r.h.s. of (\ref{9}) and the first term in the
third line of the r.h.s. of (\ref{15})
vanish. Altogether after the usual redefinition $T\rightarrow \frac{T}{2\pi}$
and the extraction of operator relations from Green functions one gets with
(\ref{5}), (\ref{9}), (\ref{15}), (\ref{19})
\bea
T_{++}(w)\Big(T_{++}(v)&+&T_{--}(v)\Big)~=~\frac{c}{2}\Big(\frac{1}{(w-v)^{4}}
+\frac{1}
{(w-\bar{v})^{4}}\Big)-\frac{\pi c}{48}\Delta \delta ^{(2)}(w-v) \nonumber \\
&+&2\Big( \frac{1}{(w-v)^{2}}+\frac{1}{(w-\bar{v})^{2}} \Big)T_{++}(w)+
\Big( \frac{1}{v-w}+\frac{1}{\bar{v}-w} \Big) \partial _{+}T_{++}(w) \nonumber
\\
&+&\Big( \frac{1}{\bar{v}-\bar{w}}+\frac{1}{v-\bar{w}} \Big) \partial _{-}
T_{++}(w)+2\pi \delta ^{(2)}(w-v) T_{+-}(w) \nonumber \\
&+& (c-\hat{c} ) \Big[~\frac{i}{12} \delta (w^{2}) \Big(\frac{2}{(w^{1}-v)
^{3}}+\frac{2}{(w^{1}-\bar{v})^{3}}+\pi \delta '(w^{1}-v^{1}) \delta (v^{2})
\Big) \nonumber \\
&&~~~~~~~~-~\frac{1}{24} \delta '(w^{2}) \Big( \frac{1}{(w^{1}-v)^{2}}
+\frac{1}{(w^{1}-\bar{v})^{2}} \Big) \Big]+ reg.~terms~.
\label{20}
\eea
This is our main result. In most cases one is interested in on shell relations
only, then $\partial _{-}T_{++}=T_{+-}=0$ and contact terms can be dropped.
Therefore, from (\ref{20}) we find $on~shell~and~for~c=\hat{c}$
\bea
T_{++}(w)\Big(T_{++}(v)+T_{--}(v)\Big)&=&\frac{c}{2}\Big(\frac{1}{(w-v)^{4}}
+\frac{1}
{(w-\bar{v})^{4}}\Big)+2\Big(\frac{1}{(w-v)^{2}}+\frac{1}
{(w-\bar{v})^{2}}\Big)T_{++} \nonumber \\
&+& \Big(\frac{1}{v-w}+\frac{1}{\bar{v}-w}\Big)\partial_{+}T_{++}(w)+reg.~
terms~~.
\label{21}
\eea
For the line of arguments presented in the introduction it is crucial, that
the factor multiplying $(c-\hat{c})$ in eq. (\ref{20}) contains also
non-contact terms.\\
Eq. (\ref{21}) is just what one usually gets for $T_{++}T_{++}+T_{++}T_{--}$
making use of the continuation in the whole 2D plane via $T_{--}(z)=
T_{++}(\bar{z})$. The last property is a consequence of analyticity and
$T_{++}=T_{--}~on~the~boundary$. Within our approach we have no thorough
method to disentangle the two products, but it seems naturally that the
terms containing $v$ or $\bar{v}$
are due to the first or second product, respectively. There is necessary
an investigation of the tensor structure of $T_{ab}T_{cd}$ using
only general covariance and $T_{a}^{a}=0$.
\section{Concluding remarks}
As anticipated in the introduction the main conclusion of our analysis is that
the standard $T\cdot T$ - OPE for the half plane can be derived without making
use of the condition of vanishing energy-momentum flux across the boundary.
All one needs is the Weyl anomaly. Thus our discussion illustrates
a further aspect of the fact that in two dimensions boundary critical behaviour
is determined by the bulk properties only \cite{C3,CL}.\\

In the $\sigma $-model picture of open
strings in general background fields one obtains equations of motion for these
background fields by the requirement of vanishing $T^{a}_{a}$ in a
distributional sense. This means $\bar{\beta}_{i}=0$ in the bulk for
fields $i$ coupling in the bulk and $\bar{\beta}_{j}=0$ on the
boundary for fields $j$ coupling only on the boundary. As argued
in ref. \cite{BD2} the finite jump of $\bar{\beta}_{i}$ in going from the
bulk to the boundary is irrelevant and yields no unwanted additional
condition for the background fields. What concerns equations for these target
space fields, the role of (\ref{3}) is a different one. While the $\bar{\beta}
_{i}=0$ conditions refer to the target space only, condition (\ref{3})
necessarily contains 2D derivatives of the string position. It is a (boundary)
condition on the 2D surface. Therefore, even if beyond the $T\cdot T$- OPE
problem the condition (\ref{3}) should remain an independent one, there would
arise no additional equation for the target space fields.
\vspace{5mm}
\\
{ \bf Acknowledgement} \\
We thank K. Behrndt, S. F\"orste and H.-J. Otto for useful discussions.
\vspace{10mm}
\\
{\bf Note added in proof} \\
The calculations of the present paper are based on the diploma thesis of
V. Preu\ss \\
(HU Berlin, February 1993).\\
H. D. thanks H. Osborn for very interesting discussions during the
Wendisch-Rietz
symposium as well as the
information that ref.\cite{AO} contains a derivation of $T_{12}=0$ from
diffeomorphism and Weyl invariance. Our energy-momentum tensor differs from
that in ref.\cite{AO}. It includes the terms which are treated there as
local boundary operators.

\end{document}